\documentclass[reprint,superscriptaddress,showpacs,amsmath,amssymb,aps,pra,lengthcheck,]{revtex4-1}
\pdfoutput=1
\usepackage{mathptmx}
\usepackage{mathtools}         
\usepackage[squaren]{SIunits}  
\usepackage{graphicx}    
\usepackage{color}          
\usepackage{hyperref}
\newcommand{\bra}[1]{\langle #1|}
\newcommand{\ket}[1]{|#1\nolinebreak[4]\rangle}

\newcommand{\abs}[1]{\left\lvert #1 \right\rvert}

\DeclareMathOperator{\Tr}{Tr}
\DeclareMathOperator{\sgn}{sgn}

\usepackage{amsthm}
\usepackage{amssymb}                         
\newtheorem*{lem}{Lemma}

\begin{document}

\title{Creating nuclear spin entanglement using an optical degree of freedom}

\author{Marcus Schaffry}
\affiliation{Department of Materials, Oxford University, Oxford OX1 3PH, United Kingdom}
\author{Brendon W. Lovett}
\affiliation{School of Engineering and Physical Sciences, Heriot Watt University, Edinburgh EH14 4AS, United Kingdom}
\affiliation{Department of Materials, Oxford University, Oxford OX1 3PH, United Kingdom}
\author{Erik M. Gauger}
\affiliation{Department of Materials, Oxford University, Oxford OX1 3PH, United Kingdom}

\date{\today}

\begin{abstract}
Molecular nanostructures are promising building blocks for future quantum technologies, provided methods of harnessing their multiple degrees of freedom can be identified and implemented. Due to low decoherence rates nuclear spins are considered ideal candidates for storing quantum information while optical excitations can give rise to fast and controllable interactions for information processing. A recent paper (Physical Review Letters \textbf{104} 200501) proposed a method for entangling two nuclear spins through their mutual coupling to a transient optically excited electron spin. Building on the same idea, we here present an extended and much more detailed theoretical framework, showing that this method is in fact applicable to a much wider class of molecular structures than previously discussed in the original proposal.
\end{abstract}

\maketitle

\section{Introduction\label{sec:introduction}}

The controlled generation of entanglement is a crucial task in quantum computing, quantum cryptography, quantum error correction and for other quantum technologies. Experimentally, controlled entangling operations have been demonstrated in a wide range of systems, such as for example pairs of photons \cite{Ou.Mandel1988ViolationofBells}, atoms \cite{Hagley.Maitre.ea1997GenerationofEinstein-Podolsky-Rosen}, ions \cite{Rowe.Kielpinski.ea2001Experimentalviolationof}, as well as between an atom and a photon \cite{Wilk.Webster.ea2007Single-AtomSingle-PhotonQuantum}, an ion and a photon \cite{Blinov.Moehring.ea2004Observationofentanglement}, among many others. However, a different approach typically needs to be used for each of these systems, so that a given control methods may not readily be transferred to a different physical system.

Employing molecular systems as `quantum hardware' \cite{Stobinska.Milburn.ea2009ScalableQuantumComputing,Benjamin.ea2006.TowardsAFullereneBasedQC} offers the advantages of high reproducibility, chemically engineered system properties, and the potential for self-assembly into more complex functional units.  The key challenge for nuclear spin-spin entanglement is achieving fast and switchable control over the interactions between adjacent quantum bits (qubits). 
Several previous publications have proposed introducing a mediator spin, whose mutual coupling to the qubit spins provides a route for a controlled entangling operation \cite{stoneham03,Benjamin.Bose2003QuantumComputingwith,Benjamin.Lovett.ea2004Opticalquantumcomputation,Ashhab.Niskanen.ea2008Interqubitcouplingmediated,Gauger.Rohde.ea2008Strategiesentanglingremote}. In these studies the mediator spin is usually of the same type as the qubit spins and possesses spin 1/2. In contrast, many molecular structures possess optically-excited triplet states (i.e. states with spin 1 character) that could be used as mediators for nearby nuclear spins \cite{Yago.Link.ea2007Pulsedelectronnuclear,*Sloop.Yu.ea1981Electronspinechoes,*Donckers.Schwencke.ea1992electronnucleardoubleresonance,*Zimmermann.Schwoerer.ea1975Endoroftriplet,*Berg.Heuvel.ea1998PulsedENDORStudies}. Motivated by recent experiments \cite{Schaffry.Filidou.ea2010EntanglingRemoteNuclear,Filidou.others} we develop theoretical control strategies for this class of systems.

\section{Model}
\begin{figure}[ht!]
  \centering
  \includegraphics{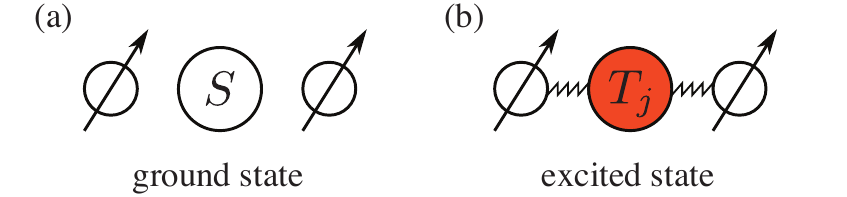}
\caption{(Color online) (a) There is no coupling between the mediator and the two nuclear spins in the ground (or vacuum) state $|0\rangle$. (b) The excited state of the mediator (e.g.~an electron-hole pair) has spin 1 character and couples to both nuclear spins. 
}
\label{fig:schema1}
\end{figure}

For our model we consider a structure comprising three integral components: two nuclear spin qubits labeled $n$ and $n'$ and a mediator spin system with an optical degree of freedom. We denote the ground state of the mediator as $\ket{0}$ and the first excited state as $\ket{e}$. Further, $\ket{0}$ is assumed to be spin-silent, while $\ket{e}$ possesses an electronic spin 1 character. The spin 1 (quasi)particle is thus created upon optical excitation of the mediator system. Importantly, the two spin qubits do not directly interact with each other, yet both are coupled to the excited state of the mediator with an isotropic Heisenberg interaction as schematically depicted in Fig.~\ref{fig:schema1}.
The Hamiltonian for such a three particle system in an external magnetic field is then given by
\begin{multline}
      \label{eq:1}
H=-\omega_n S_{z,n}- \omega_{n'} S_{z,n'}+\ket{e} \bigl( \omega_e S_{z,e} +\omega_0 \bigr) \bra{e} \\
+\ket{e} \bigl( A  \mathbf{S}_n\cdot \mathbf{S}_e + A' \mathbf{S}_{n'}\cdot \mathbf{S}_e   + D S_{z,e}^2 \bigr) \bra{e}\quad,  
\end{multline}
where $\omega_{n/n'}$ denotes the Zeeman splitting of the two qubits and $\omega_e$ that of the central spin 1 particle; $D$ is the uniaxial zero-field-splitting, $A$ and $A'$ are the isotropic Heisenberg coupling constants, and $\omega_0$ denotes the (typically optical) excitation energy of the mediator. Here $S_{z,n/n'/e}$ and $\mathbf{S}_{n/n'/e}$ are the usual component and total vector Pauli spin operators respectively. Throughout this article we assume $\abs{D},~A \ll \omega_e$ and $\omega_e>0$.

\begin{figure}[ht!]
  \centering
  \includegraphics[width=0.8\linewidth]{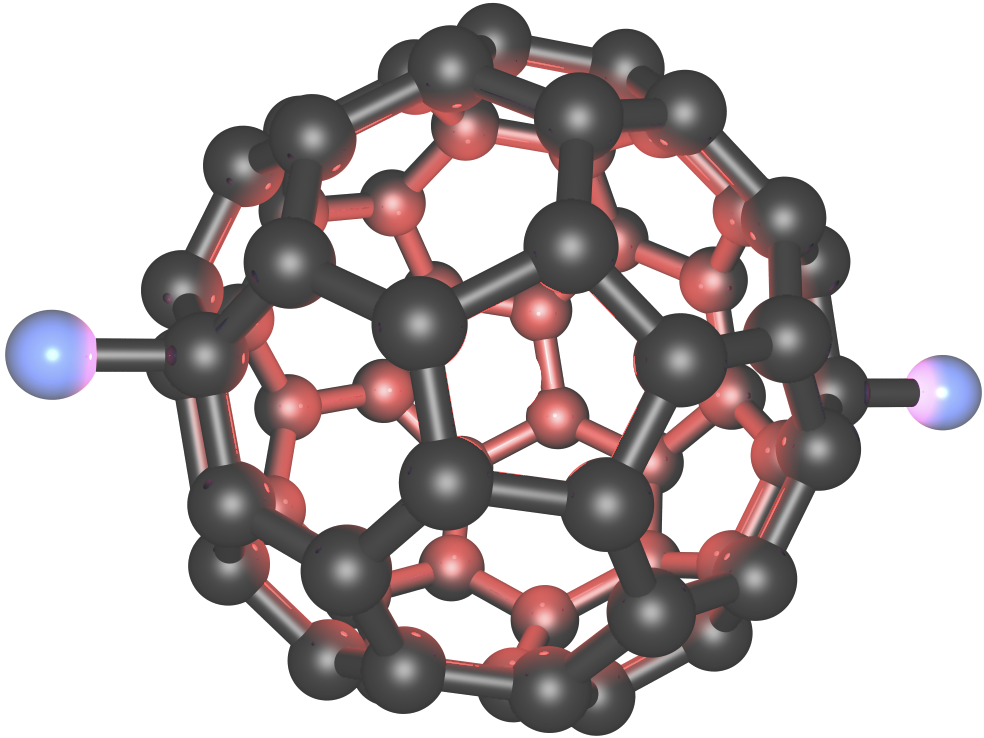}
    \caption{(Color online) A functionalised buckyball with two satellite atoms (blue). In reality the functional groups are likely much larger than a single atom, but for simplicity we here only depict the two relevant atoms with a nuclear spin 1/2, for example $^{13}$C, $^{15}$N, or $^{31}$P. Note that the two satellites (or rather functional groups) are not necessarily both the same, and that the system may thus be `asymmetric'.}
  \label{fig:schema3}
\end{figure}

The Hamiltonian \eqref{eq:1} can describe many different nanostructures, in particular a range of optically active molecules, in which case the mediator ground (excited) state corresponds to the absence (presence) of an electron-hole pair. The transition between these two states can be induced by a short laser pulse of frequency $\omega_0$; moreover, the excited state will decay naturally through spontaneous emission. By contrast, an NV$^-$-centre in diamond surrounded by two $^{13}$C atoms \cite{Neumann.Mizuochi.ea2008MultipartiteEntanglementAmong} is an example of a system in which the spin 1 mediator is not susceptible to decay and is ever-present. Many of the results discussed in this article are equally valid for this kind of system. However, the following discussion primarily focuses on the example case of the molecular system consisting of a functionalised C$_{60}$ molecule with two functional groups as depicted in Fig.~\ref{fig:schema3}. 

In what follows we will first we analyze Hamiltonian \eqref{eq:1} for two different cases: that of a symmetric and that of an asymmetric system. In order to understand the system properties and dynamics, we will determine the eigensystem using perturbation theory. We next present suitable protocols for the controlled generation of nuclear spin entanglement for both cases. We conclude the paper with a brief summary and a discussion of our results.

\section{Eigenspectrum and effective Hamiltonian}
\label{sec:eigenspectras}

\subsection{Symmetric system}
\label{sec:symmetric-case}
By definition the symmetric system consists of two nuclear spins with $\omega_n=\omega_{n'}$ and equal hyperfine coupling constants $A = A'$. In the presence of the electronic excitation, i.e.~after the application of a suitable laser pulse, the Hamiltonian governing the spin dynamics is given by
\begin{multline}
    H_{\text{sym}}= \bra{e}H\ket{e}=-\omega_n S_{z,n}+\omega_e S_{z,e} - \omega_n S_{z,n'} \\+ A \left( \mathbf{S}_n\cdot \mathbf{S}_e + \mathbf{S}_{n'}\cdot \mathbf{S}_e \right)  + D S_{z,e}^2 \quad,
\end{multline}
where we have neglected the term $\omega_0$ that is proportional to the identity and unimportant for the dynamics. Since the electronic Zeeman splitting $\omega_e$ is typically the largest energy scale of the system (and in particular much larger than $\omega_n$), it is safe to assume that
\begin{equation}
  \label{eq:assum}
\abs{\omega_n},\abs{D},A\ll \omega_e\quad.
\end{equation}

Based on the above assumption, we employ degenerate perturbation theory to determine the eigenspectrum of $H_{\text{sym}}$. We begin by partitioning the Hamiltonian as follows
\begin{equation}
  H_{\text{sym}}= H_{0,\text{sym}} + H_{\text{sym}}' \quad,
\end{equation}
where
\begin{equation}
H_{0,\text{sym}} = -\omega_n S_{z,n}+\omega_e S_{z,e} - \omega_n S_{z,n'} + D S_{z,e}^2  
\end{equation}
will be treated exactly, and the perturbation is given by
\begin{equation}
  H_{\text{sym}}' = A \left( \mathbf{S}_n\cdot \mathbf{S}_e + \mathbf{S}_{n'}\cdot \mathbf{S}_e \right)  \quad.
\end{equation}

\begin{figure}[ht!]
  \centering
  \includegraphics{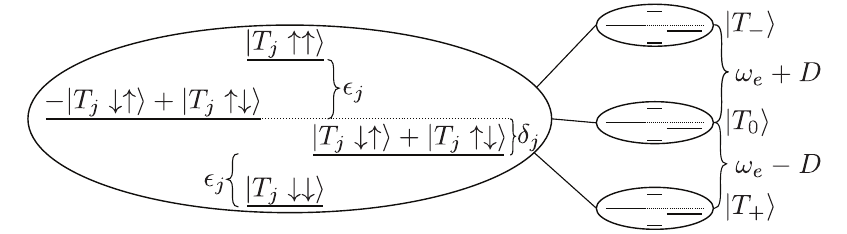}
  \caption{Eigenspectrum of a symmetric system up to second order of perturbation theory (eigenvectors not normalized). Our basis states obey the following relations $S_{z,e}|T_{\pm}\rangle = \mp |T_{\pm} \rangle$, $S_{z,e}|T_{0}\rangle = 0$, $S_{z,n} |\downarrow \rangle = |\downarrow \rangle$, and $S_{z,n} |\uparrow \rangle =- |\uparrow \rangle$.\label{fig:eigen-sym}}
\end{figure}

As $H_{0,\text{sym}}$ is diagonal in the computational basis (defined as $\{\ket{en_1n_2}~|~e=T_{\pm },T_0 \text{ and } n_{1/2}=\uparrow,\downarrow\}$) and $\bra{\downarrow \uparrow}H_{0,\text{sym}}\ket{\downarrow \uparrow}=\bra{\uparrow\downarrow}H_{0,\text{sym}}\ket{\uparrow\downarrow}$, degenerate perturbation theory is required. However, since $\bra{T_j Y}H_{\text{sym}}'\ket{T_l Z}=0$ for $j,l = -,0,+$ and $Y,Z=\downarrow\uparrow, \uparrow \downarrow$  the degeneracy is not removed in the first order. To find the correct zeroth order wavefunctions, we need to solve the corresponding secular equations to second order \cite{Schiff1949QuantumMechanicsVolume1}.
Making use of the assumption \eqref{eq:assum} we then get the following second order eigenenergies:
\begin{align}
  E_{j,1} &= E_{j,3} - \epsilon_j &   E_{j,2} &= -j\omega_e +\abs{j} D\\
  E_{j,3} &= E_{j,2} - \delta_j  &   E_{j,4} &= E_{j,2} + \epsilon_j ,
\end{align}
with the corresponding eigenvectors up to first order (see Fig.~\ref{fig:eigen-sym}):
\begin{align}
  \ket{E_{j,1}} &= \ket{T_j \downarrow \downarrow} & \ket{E_{j,2}} &= \frac{1}{\sqrt{2}} \Bigl(-\ket{T_j\downarrow\uparrow} + \ket{T_j\uparrow \downarrow}\Bigr) \\
\ket{E_{j,4}} &= \ket{T_j \uparrow \uparrow} &  \ket{E_{j,3}} &=\frac{1}{\sqrt{2}} \Bigl(\ket{T_j\downarrow\uparrow} + \ket{T_j\uparrow \downarrow} \Bigr)
\end{align}
and where we have defined the following quantities for compactness:
\begin{align}
  \epsilon_{-} &= \omega_n -A +2a_- , \; \epsilon_0 = \omega_n + 2 a_+ ,\; \epsilon_{+} = \omega_n +A \quad, \\
 \delta_- &= -2 a_- ,\;  \delta_{0} = -2a_0,  \quad\text{and} \quad \delta_+ = 2 a_+ \quad, \\
   a_{\pm} &= \frac{1}{2}\frac{A^2}{\mp D + \omega_e + \omega_n} \quad\text{and} \quad a_0 = a_+ - a_- .\quad, \label{eq:defA} 
\end{align}

Based on this spectrum, we now approximate $H_{\text{sym}}$ in the following way to obtain an effective Hamiltonian:
\begin{equation}
\label{eq:appr-effe}
   H_{\text{sym}} \approx V^\dagger \text{diag}(E_{-1,1},\ldots,E_{1,4}) V =: H_{\text{sym, eff}} \quad,
\end{equation}
where $V$ is the matrix of the approximate eigenvectors (see Fig.~\ref{fig:eigen-sym})
\begin{equation}
V =  ( \ket{E_{-,1}} \ket{E_{-,2}} \cdots \ket{E_{+,4}} )  \quad.
\end{equation}
 
In the computational basis $H_{\text{sym, eff}}$ consists of three blocks with the following structure:
\begin{equation}
  \begin{pmatrix}
  \times \\
    &\times&\times \\
    &\times&\times \\
    & & & \times 
\end{pmatrix}\quad; \quad \text{$\times$ denotes a nonzero entry} \quad,
\end{equation}
where each block belongs to one of the three spin states $T_{-}$, $T_0$, or $T_{+}$. Correspondingly, we define the three subspaces $\mathcal{T}_j$ spanned by the vectors $\ket{T_j\downarrow\downarrow},\ket{T_j\downarrow\uparrow},\ket{T_j\uparrow\downarrow},$ and $\ket{T_j\uparrow\uparrow}$ with $j=-,0,+$.  this paper, the index $j$ will denote the electronic spin state of the excitation whereas the letter $i$ always indexes the four nuclear spin states. 

In writing Eq.~\eqref{eq:appr-effe} we have neglected all matrix elements between the subspaces $\mathcal{T}_i$ and $\mathcal{T}_j$ ($i\neq j$); the effect of these matrix elements is negligible because the electronic Zeeman splitting $\omega_{e}$ is much larger than the nuclear Zeeman splittings. The dynamics in each of the $\mathcal{T}_j$ is therefore closed and can be described by an effective $4 \times 4$ Hamiltonian of the form given above. We shall now analyze the nuclear spin dynamics arising from each of those effective subspace Hamiltonians. A term in the Hamiltonian that is proportional to the identity has no effect on the dynamics within the subspace, so we can subtract $D+\omega_e+a_-,~a_0,$ and $D-\omega_e-a_+$ from the subspace Hamiltonians for $\mathcal{T}_{-},~\mathcal{T}_0,$ and $\mathcal{T}_+$ respectively (thereby neglecting unimportant phases that are common to all states in each block). We can simplify the resulting Hamiltonians $H_{\text{sym, eff, }j}$ further by transforming to a suitably chosen rotating frame through the transformation
\begin{equation}
   H_{\text{sym, eff, }j}' = U_j H_{\text{sym, eff, }j} U^\dagger_j  + i U_j \frac{d U^\dagger_j}{dt}\quad,
\end{equation}
where
\begin{equation}
 U_j(t) = R_{z,n}\bigl(\phi_j t \bigr) \otimes R_{z,n'}\bigl( \phi_j t \bigr)   
\end{equation}
and
\begin{align}
\phi_{\pm} &= -\omega_n \mp A -a_{\pm}  \quad,  \label{eq:phipm} \\
\phi_0 & = -\omega_n - a_- -a_+ \quad, \label{eq:phi0}\\
R_{z,i}( \phi) &= \exp ( -i S_{z,i} \phi ) \quad .
\end{align}
In this rotating frame we  obtain the following Hamiltonians
\begin{equation}
  H_{\text{sym, eff, }j}' = k(j) 2 a_j(S_{x,n}  S_{x,n'} + S_{y,n} S_{y,n'}) \;,
\end{equation}
where $k(j) = 1,1,-1$ (with $j=-,0,+$, as it will be throughout the paper). Each of these effective subspace Hamiltonians therefore induces the dynamics of a direct XY-coupling of the two nuclear spins, with a time evolution corresponding to Rabi flopping between the nuclear spin states $\ket{\downarrow\uparrow}$ and $\ket{\uparrow\downarrow}$. The Rabi frequency $a_j$ depends on the spin state $j$ of the optical excitation. According to Eq.~\eqref{eq:defA}, we expect $a_{+}$ to be fairly similar to $a_{-}$, whereas in comparison $a_0$ will be much smaller.

We will return to the symmetric case to discuss entanglement generation in Section~\ref{sec:symmtimeev}, but first complete our analysis of the Hamiltonian for non-symmetric cases.

\subsection{Asymmetric system}
\label{sec:asymmetric-case}
Next we consider an asymmetric system with unequal nuclear Zeeman splittings and/or unequal nuclear - electronic excitation hyperfine coupling. The excited state Hamiltonian is then given by:
\begin{multline}
  \label{eq:asymmetric}
 H_{\text{asym}}=\bra{e}H\ket{e}=-\omega_n S_{z,n}+\omega_e S_{z,e} - \omega_{n'} S_{z,n'} \\
+ A  \mathbf{S}_n \cdot \mathbf{S}_e + A'\mathbf{S}_{n'}\cdot \mathbf{S}_e   + D S_{z,e}^2 \quad.
\end{multline}
For the purpose of applying perturbation theory, we consider the nuclear-spin-mediator coupling as the perturbation and split the Hamiltonian as follows
\begin{align}
H_{0,\text{asym}} &= -\omega_n S_{z,n}+\omega_e S_{z,e} - \omega_{n'} S_{z,n'} + D S_{z,e}^2 \quad \text{and} \\  
H_{\text{asym}}' &= A \mathbf{S}_n \cdot \mathbf{S}_e + A' \mathbf{S}_{n'}\cdot \mathbf{S}_e \quad.
\end{align}
For the asymmetric system we shall assume that the parameters $A$ and $A'$ and $\omega_n$ and $\omega_{n'}$ differ sufficiently to fulfil the following two inequalities
\begin{align}
   \label{eq:cond}
\frac{1}{2}\abs{A'-A \pm (\omega_{n'}-\omega_n)} &\gg \abs{a_{\pm}},\abs{a_{\pm}'} \quad, \\
\frac{1}{2}\abs{\omega_{n'}-\omega_n} &\gg \abs{a_0}, \abs{a_0'} \quad, \label{eq:condB}
\end{align}
where $a_{j}$ and $a_{j}'$ are as defined in the previous section, with the latter using $\omega_{n'}$ and $A'$ rather than $\omega_{n}$ and $A$. Under these assumptions, non-degenerate perturbation theory can be used for calculating the eigensystem. The state corrections to first order perturbation theory then introduce a small mixing of computational basis states with corrections of magnitude 
\begin{equation}
\label{eq:corrections}
   \frac{1}{\sqrt{2}}\frac{A}{D \pm( \omega_e + \omega_n)} \text{  and } \frac{1}{\sqrt{2}}\frac{A' }{D \pm( \omega_e + \omega_{n'})}  \quad,
\end{equation}
which are negligible for $A,A' \ll \omega_e$. The eigenstates of $H_{\text{asym}}$ thus coincide with those of $H_{0,\text{asym}}$ (which are simply the computational basis states) to a very good approximation. 

Analogously to our approach in the symmetric case, we proceed by analyzing individual effective Hamiltonians for the subspaces $\mathcal{T}_{j}$  ($j=-,0,+$). After adding a suitable constant to each subspace, we obtain three diagonal Hamiltonians
\begin{equation}
\label{eq:AsymmetricHamiltonian}
  H_{\text{asym, eff,}  j}= \phi_{j}'  S_{z,n'} + \phi_{j} S_{z,n} \quad ,
\end{equation}
where $\phi_j', \phi_j$ are as defined in Eqs.~\eqref{eq:phipm} and \eqref{eq:phi0} and the prime denotes that $\omega_{n'}$ and $A'$ are used for the expression instead of $\omega_n$ and $A$. All three effective Hamiltonians are local, meaning there is no direct coupling between the two nuclear spins.

\subsection{Crossover between a symmetric and an asymmetric system}
\label{sec:cross-betw-symm}
In a certain region of parameter space neither the symmetric nor the asymmetric analyses are justified. In this section we consider this `crossover' regime, so that in the later discussion we will be able to interpolate between those two cases.

We start with the approximated symmetric Hamiltonian $H_{\text{sym,eff}}$ and treat the asymmetry as the perturbation $H_{\text{co}}'$, i.e.\ 
\begin{align}
H_{\text{co}} &= H_{\text{sym,eff}} +H_{\text{co}}' \quad, \\
H_{\text{co}}' &= -\Delta_1\omega_{n} S_{z,n'} + \Delta_2 A \mathbf{S}_{n'} \cdot \mathbf{S}_e   \quad,
\end{align}
where $\Delta_1$ is the fractional difference between the two nuclear Zeeman splittings, $\Delta_1= \frac{\omega_{n'} -\omega_{n}}{\omega_n}$, and $\Delta_2$ is the fractional difference between the two coupling constants, $\Delta_2=\frac{A'-A}{A}$.

We begin with the vectors $\ket{E_{j,i}}$ and values $E_{j,i}$  ($j=-,0,+; i=1,\dots,4$) from section \ref{sec:symmetric-case} as the eigenstates and eigenvalues of Hamiltonian $H_{\text{sym,eff}}$. As in the other two cases, we can still neglect those terms of $H_{\text{co}}$ which couple different projections of the excitation spin with strength $\Delta_2 K/\sqrt{2}$ since the relevant states differ in energy by about $\omega_e$, yielding an effective crossover Hamiltonian $H_{\text{co,eff}}$ with the eigenvectors:
\begin{align}
  \ket{\widetilde{E}_{j,1}} &= \ket{T_j \downarrow \downarrow}; & \ket{\widetilde{E}_{j,2}} &= \frac{1}{\sqrt{2}} \Bigl(\alpha_{j,2,1}\ket{T_j\downarrow\uparrow} + \alpha_{j,2,2} \ket{T_j\uparrow \downarrow}\Bigr); \\
\ket{\widetilde{E}_{j,4}} &= \ket{T_j \uparrow \uparrow}; &  \ket{\widetilde{E}_{j,3}} &=\frac{1}{\sqrt{2}} \Bigl(\alpha_{j,3,1} \ket{T_j\downarrow\uparrow} + \alpha_{j,3,2}\ket{T_j\uparrow \downarrow} \Bigr);
\end{align}
where
\begin{equation}
\label{eq:alpha}
  \abs{ \alpha_{j,2,1/2}}^2 =  \abs{\alpha_{j,3,2/1}}^2 = \frac{1}{2} \left( 1 \pm \frac{\sgn(a_j)f_j}{\sqrt{a_j^2 +f_j^2}} \right) 
\end{equation}
and 
\begin{equation}
\label{eq:defOfF}
f_{\pm} = \frac{1}{2} (\Delta_2 A \pm \Delta_1 \omega_n) \quad \text{and}  \quad   f_0 = -\frac{1}{2}\Delta_1 \omega_n \quad .
\end{equation}
The corresponding eigenvalues are
\begin{align}
  \widetilde{E}_{j,1/4} &=  E_{j,1/4} \pm k(j) f_j \\
 \widetilde{E}_{j,2/3} &= E_{j,2} + k(j) \left(a_j \mp \sgn(a_j)\sqrt{a_j^2 +f_j^2} \right) \quad,
\end{align}
where $k(j)$ is again $1,1,-1$ for $j=-,0,+$ respectively. 

It is easy to see that the eigenstates reduce to those of the symmetric case for a vanishing  perturbation ($f_j=0$), whereas they tend to the computational basis states for a larger perturbation as required for the  asymmetric system ($|f_j| \gg \abs{a_j}$). 

\section{Controlled generation of entanglement}
The contrasting analyses presented in the previous section suggest that the dynamics of the system will vary significantly depending on its parameters. Therefore, different approaches for achieving controlled entanglement between the two nuclear spins will be required. In this section we will show how to do this in each case; we begin with the symmetric system introduced in section~\ref{sec:symmetric-case}. 

\subsection{Symmetric system: entangling time evolution}
\label{sec:symmtimeev}
For the symmetric system, we shall exploit the effective XY-coupling between the spin states $\ket{\uparrow\downarrow}$ and $\ket{\downarrow\uparrow}$ for the generation of entanglement. The fact that the magnitude of this coupling depends on the spin state of the excitation will allow us to control the interaction.

The free time evolution in any of the subspaces $\mathcal{T}_j$ takes suitable initial product states of the nuclear spins to entangled states at certain subsequent points of time. In order to quantify the performance of the desired operation, we consider the \textit{entangling power} of the unitary operator that describes the time evolution of the system. The entangling power is defined as the mean linear entropy produced by the unitary operator acting on a uniform distribution of all (pure) product states \cite{Zanardi.Zalka.ea2000Entanglingpowerof}. A maximally entangling two qubit gate, e.g.\ the CNOT- or the CPHASE-gate, possesses an entangling power of $2/9$.

\begin{figure}[h!]
  \centering
  \includegraphics{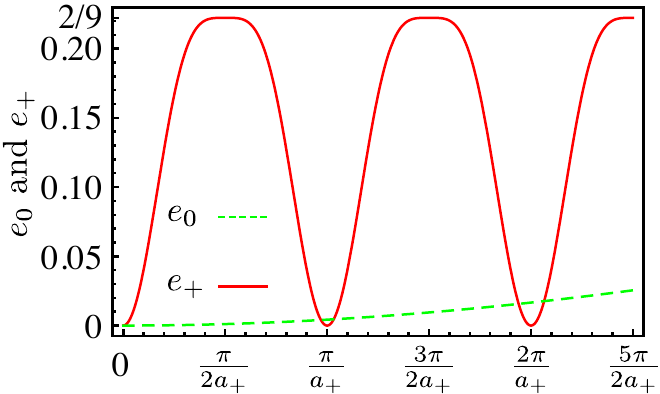}
  \caption{(Color online) Entangling power $e_{+/0}$ of the time evolution operators $U_{+/0}(t)$. Here $a_+=32 a_0$ (see text).}
\label{fig:entangling-power}
\end{figure}

For the symmetric system the entangling power of the free time evolution $U_j(t)=\exp(-i H_{\text{sym,eff,}j}' \,t)$ is given by the following simple expression for each of the three subspaces $\mathcal{T}_{j}$:
\begin{equation}
e_{j}=\frac{1}{9} \bigl(3 + \cos(2a_j t) \bigr) \sin^2(a_j t) \quad.
\end{equation}
The entangling power $e_j$ is only a function of the coupling strength $a_j$ between the states $\ket{T_j\downarrow\uparrow}$ and $\ket{T_j \uparrow\downarrow}$, becoming maximally entangling at times that are odd integer multiples of $t_{\rm max}={\pi}/{(2a_j)}$. However, the characteristic timescale over which entanglement builds up differs significantly between the subspaces, since ${a_{\pm}}/{a_0} = {\omega_e}/{(2D)}$ to leading order, and this ratio is typically much larger than one. Hence, $e_0(t)\approx 0$ for $t< {\pi}/{(2a_{\pm})}$ when $\abs{D} \ll \omega_e$ as we have assumed so far. Fig.~\ref{fig:entangling-power} shows the entangling powers $e_+$ and $e_0$ for a typical ratio of ${a_+}/{a_0}=32$.

Based on these observations, we can formulate a protocol for the controlled generation of entanglement. Consider a system that is initially in its ground state (i.e. there is no excitation coupling to the nuclear spins). A laser pulse creates the excitation, which will in general be in a mixture of the three spin states $\ket{T_-},\ket{T_0},$ and $\ket{T_+}$. For the present discussion, we will assume it to be in the state $\ket{T_0}$; the discussion of an mixed initial state will follow later. A short microwave pulse allows us to flip the state of the excitation selectively to either the $\ket{T_+}$ or the $\ket{T_-}$ state since the transition frequencies are split by the ZFS. The entangling dynamics proceeds much faster in these outer subspaces, reaching the first entangling power maximum after a time $t=\frac{\pi}{2a_{\pm}}$. We can now apply another microwave pulse to flip the excitation back into the $\ket{T_{0}}$ state, where the further evolution is much slower and entanglement can be preserved.

There are, however, a few subtle points worth pointing out. First, similar to the case of applying the maximally entangling CNOT operation, only suitable initial states will evolve into entangled states. Second, certain particular input states can reach maximal entanglement in less time than $\frac{\pi}{2a_{\pm}}$; for example, the time evolution $U(\frac{\pi}{4a_+})$ takes $\ket{T_+\downarrow\uparrow}$ to a maximally entangled state. 

Finally, the (optical) excitation, whose natural lifetime $\tau$ is assumed to be longer than the time it takes to build up the entanglement $\frac{\pi}{2a_{\pm}}<\tau$, needs to be destroyed quickly enough such that no further (slower) evolution unwinding the achieved entanglement occurs in the $\mathcal{T}_0$ subspace. This can either be achieved with a coherent optical $\pi$-pulse, or alternatively by simply waiting for the system to decay back to its ground state if the following hierarchy of timescales exists in the system
\begin{equation}
\label{eq:inequality-tau}
 \frac{\pi}{2a_{\pm}} < \tau \ll \frac{\pi}{2\abs{a_0}}  \quad.
\end{equation}
However, in contrast to the de-excitation with a coherent laser pulse, it is not immediately obvious that the nuclear spin entanglement can survive the optical decay process; we will therefore now take a small diversion to analyse this decay process in detail.

\subsubsection*{Decoherence due to the optical decay process}
\label{sec:decoherence}
In general, the optical decay of the mediator induces decoherence on the nuclear spins. In order to quantify this, we will make use of the quantum optical master equation (for a full derivation see e.g.\ Ref.~\onlinecite{Breuer.Petruccione2002TheoryofOpen}):
\begin{multline}
  \label{eq:master-full}
\frac{d}{dt} \tilde{\rho}(t) = \sum_{\omega,\omega'} e^{i(\omega' -\omega) t} \Gamma(\omega) \Bigl( A(\omega) \tilde{\rho}(t) A(\omega')^{\dagger} \\ -A(\omega')^{\dagger}A(\omega) \tilde{\rho}(t) \Bigr) + \text{H.c.} \quad,
\end{multline}
where $\tilde{\rho}$ denotes the density matrix in the interaction picture, $\Gamma(\omega')$ is the rate for a transition with frequency $\omega$ and H.c.\ is the Hermitian conjugate. The sum is taken over all optical transitions of the system with transition operators $A(\omega)$ as defined by
\begin{equation}
\label{eq:transitionOperator}
  A(\omega) = \sum_{E'- E= \omega} \Pi(E) \mathcal{D} \Pi(E')  \quad,
\end{equation}
where $E$ and $E'$ are eigenvalues of $H$ that differ by $\omega$ and $\Pi(E)$ denotes the projection onto the eigenspace belonging to the eigenvalue $E$. $\mathcal{D}$ denotes the system's optical dipole operator. The symmetric system features twelve optical transitions $\ket{eT_j \downarrow \downarrow} \rightarrow \ket{0\downarrow \downarrow}$, $\ket{eT_j}\frac{1}{\sqrt{2}}\bigl(\mp\ket{\downarrow \uparrow}+\ket{\uparrow \downarrow}\bigr) \rightarrow \ket{0}\frac{1}{\sqrt{2}}\bigl(\mp\ket{\downarrow \uparrow}+\ket{\uparrow \downarrow}\bigr)$,  and $\ket{eT_j \downarrow \downarrow} \rightarrow \ket{0\downarrow \downarrow}$ for $j=-,0,+$. We make the additional assumption that the optical decay rates are all equal and can thus be characterised by a single optical lifetime $\tau$.

Equation \eqref{eq:master-full} can often be simplified by applying an instance of a rotating wave approximation (RWA), based on the assumption that fast oscillating terms average out \cite{Breuer.Petruccione2002TheoryofOpen}, giving this more common form of the quantum optical master equation:
\begin{multline}
  \label{eq:master-rwa}
\frac{d}{dt} \tilde{\rho}(t) = \sum_{\omega}  \Gamma(\omega)\Bigl( A(\omega) \tilde{\rho}(t) A(\omega)^{\dagger}  \\-A(\omega)^{\dagger}A(\omega) \tilde{\rho}(t) \Bigr) + \text{H.c.} \quad.
\end{multline}
However, the RWA is only justified when $\abs{\omega-\omega'}^{-1}$ is small compared to the relaxation time of the system $\tau$. Under the assumptions of equation \eqref{eq:inequality-tau} this is fulfilled except for the two transition frequencies $\omega_1=E_{0,2}+\omega_0$ and $\omega_2=E_{0,2}+\omega_0-\delta_0$ which corresponds to the transitions $\ket{eT_0}\frac{1}{\sqrt{2}}\bigl(\mp\ket{\downarrow \uparrow}+\ket{\uparrow \downarrow}\bigr) \rightarrow \ket{0}\frac{1}{\sqrt{2}}\bigl(\mp\ket{\downarrow \uparrow}+\ket{\uparrow \downarrow}\bigr)$. Therefore we can safely apply the RWA to all remaining frequencies, obtaining:
\begin{align}
  \label{eq:master-mix}
  &\begin{multlined}[t]
\frac{d}{dt} \tilde{\rho}(t) =
\sum_{\omega,\omega' \in S} e^{i(\omega' -\omega) t} \Gamma(\omega) \Bigl( A(\omega) \tilde{\rho}(t) A(\omega')^{\dagger} \\ -A(\omega')^{\dagger}A(\omega) \tilde{\rho}(t) \Bigr) \\
+\sum_{\omega \not\in S} \Gamma(\omega) \Bigl( A(\omega) \tilde{\rho}(t) A(\omega)^{\dagger}  -A(\omega)^{\dagger}A(\omega) \tilde{\rho}(t) \Bigr) + \text{H.c.}   
  \end{multlined}\\
 &\begin{multlined}[t]
= \sum_{\omega} \Gamma(\omega) \Bigl( A(\omega) \tilde{\rho}(t) A(\omega)^{\dagger}  -A(\omega)^{\dagger}A(\omega) \tilde{\rho}(t) \Bigr) \\
+ e^{2a_0 i t} \frac{1}{2\tau} \Bigl( 2 A(\omega_1) \tilde{\rho}(t) A(\omega_2)^{\dagger}-\\
 A(\omega_2)^{\dagger} A(\omega_1) \tilde{\rho}(t) - \tilde{\rho}(t) A(\omega_2)^{\dagger} A(\omega_1) \Bigr) + \text{H.c.} 
  \end{multlined}
\end{align}
where $S=\{\omega_1,\omega_2\}$ and using phenomenological decay rates
\begin{equation}
  \label{eq:rates}
  \Gamma(\omega) =
  \begin{cases}
\frac{1}{2 \tau}  & \text{ if } \omega >0  \\
0   & \text{ if } \omega < 0
  \end{cases} \quad.
\end{equation}
This definition of decay rates describes the typical situation of only spontaneous emission occurring, since stimulated emission or absorption of photons are proportional to the negligible power density of thermally activated photons in the environmental modes.
 
Rather than solving the above master equation for the entire Hilbert space of our system, we are here only interested in the $4 \times 4$ density matrix $\rho^f$ of the two nuclear spins after the decay has occurred:
\begin{equation}
 \rho^f :=  \bra{0} \tilde{\rho}(t= \infty) \ket{0} \quad. 
\end{equation}
Limiting the following discussion to this nuclear spin subspace, it is easy to see that only a small number of elements of  $\rho^f$ can be populated by the decay process. These non-zero elements are $\rho^f_{a,b}$ with $a=b=\downarrow\downarrow$, $a=b=\uparrow\uparrow$, and $a,b \in \{\downarrow\uparrow, \uparrow\downarrow \}$. The final nuclear spin state after the optical decay can be compactly written as
\begin{multline}
 \rho^f_{a,b}= \bra{eT_0a} \tilde{\rho}_0 \ket{eT_0 b}+\frac{1}{2} \sum_{j=\pm} \bigl( \bra{eT_ja}\tilde{\rho}_0 \ket{eT_j b} \\
+ \bra{eT_j\overline{a}}\tilde{\rho}_0 \ket{eT_j \overline{b}} \bigr) \quad,
\end{multline}
where $\tilde{\rho}_0=\tilde{\rho}(0)$ is the full system's density matrix in the interaction picture before the decay process and the bar on top of the nuclear spin states denotes that these are flipped, i.e.\ $\overline{\downarrow}=\uparrow$ and $\overline{\uparrow}=\downarrow$. Hence the coherences between $\ket{eT_{\pm}\downarrow\uparrow}$ and $\ket{eT_{\pm}\uparrow\downarrow}$ do not necessarily survive, but all the coherences between $\ket{eT_0 \downarrow \uparrow}$ and $\ket{eT_0 \uparrow \downarrow}$ survive the optical decay, reflecting the fact that no RWA approximation has been made for the transitions $\ket{e T_0} \bigl(\pm\ket{\downarrow \uparrow}+ \ket{\uparrow \downarrow}\bigr) \rightarrow \ket{0} \bigl(\pm\ket{\downarrow \uparrow}+ \ket{\uparrow \downarrow}\bigr)$. Physically, this means that these two transitions produce indistinguishable photons due to overlapping emission spectra. In the absence of other decoherence processes the spectra are simply given by lifetime broadened Lorentzians \cite{Mandel.Wolf1995Opticalcoherenceand}:
\begin{equation}
L({\omega}) = \frac{1/\tau}{(\omega -\omega_i)^2 + (1/\tau)^2}  \quad\text{for } i=1,2 \quad. 
\end{equation}
The photons emitted in all other ten transitions can be in principle be distinguished, so that we are effectively dealing with eleven distinct, incoherent decay channels. It is worth noting that there are three decay channels which populate the $\ket{0}\bigl(\pm \ket{\downarrow \uparrow} + \ket{\uparrow \downarrow} \bigr)$ states, one each for the excitation spin projections ${T}_{0,\pm}$, but only the decays from the $\mathcal{T}_0$ subspace preserve coherence.

\begin{figure}[ht!]
  \centering
  \includegraphics{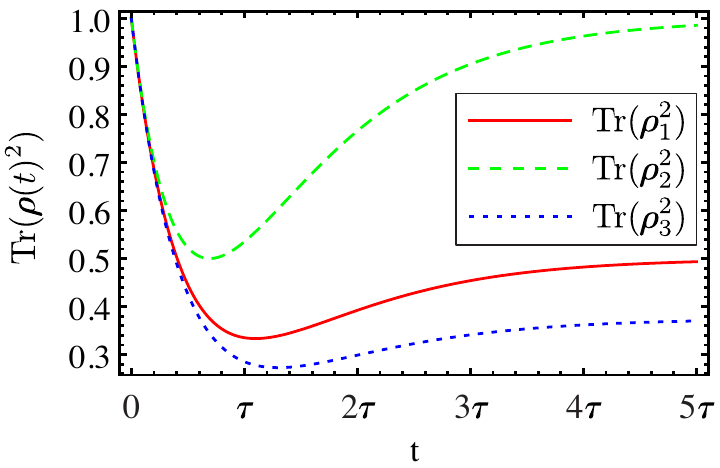}
\caption{(Color online) Evolution of the purity $\Tr(\rho(t)^2)$ of a system initialized in an excited state during the optical decay process. The three different initial states shown are defined in \eqref{eq:purity-intial-states}, and $\tau \delta_0 =0.05\ll \pi$.}
  \label{fig:purity-sym}
\end{figure}

In Fig.~\ref{fig:purity-sym} we plot the purity $\Tr(\rho(t)^2)=\Tr(\tilde{\rho}(t)^2)$ to illustrate the decoherence induced by the optical decay for three different initial density matrices $\rho_i(0)=\ket{\psi_i}\bra{\psi_i}$ with
\begin{equation}
\label{eq:purity-intial-states}
  \begin{split}
\ket{\psi_1} &= \ket{eT_+\uparrow\downarrow},  \qquad\ket{\psi_2} = \ket{eT_0\uparrow\downarrow} \quad, \quad \text{and}\\
\ket{\psi_3} &= \frac{1}{2}\ket{eT_0} \bigl( \ket{\downarrow\downarrow} + \ket{\downarrow \uparrow} + \ket{\uparrow\downarrow} +\ket{\uparrow \uparrow} \bigr) \quad.
  \end{split}
\end{equation}
Solving the master equation \eqref{eq:master-mix} analytically, we then obtain the following final density matrices of the nuclear spins after the decay:
\begin{align}
 \rho_1^f &= \frac{1}{2}\bigl( \ket{\downarrow\uparrow}\bra{\downarrow\uparrow} +\ket{\uparrow\downarrow}\bra{\uparrow\downarrow} \bigr) \quad, \\
 \rho_2^f &
 \begin{multlined}[t]
= \frac{1}{2+2\delta_0^2\tau^2} \Bigl( \delta_0^2\tau^2 \ket{\downarrow\uparrow}\bra{\downarrow\uparrow} + i \delta_0\tau  \ket{\downarrow \uparrow}\bra{\uparrow\downarrow}  \\
- i \delta_0\tau  \ket{\uparrow\downarrow} \bra{\downarrow\uparrow}+ (2+\delta_0^2\tau^2 ) \ket{\uparrow\downarrow}\bra{\uparrow\downarrow} \Bigr) \quad ,
 \end{multlined}\\
\rho_3^f &= \frac{1}{4} \Bigl( \mathbf{1} + \ket{\downarrow \uparrow}\bra{\uparrow\downarrow}+ \ket{\uparrow\downarrow} \bra{\downarrow\uparrow} \Bigr) \quad ,
\end{align}
with the corresponding purities
\begin{align}
 \Tr(\rho_1^f)^2 &=1/2 \quad \text{and} \quad \Tr(\rho_3^f)^2 =3/8 \quad,\\
 \Tr(\rho_2^f)^2 &=\frac{2+\delta_0^2\tau^2}{2+2 \delta_0^2\tau^2} \overset{\eqref{eq:inequality-tau}}{\approx} 1 -\frac{\delta_0^2\tau^2}{2}\quad.
\end{align}

At first it may seem surprising that an initial state $\ket{eT_+}\ket{\uparrow\downarrow}$ can end up in a complete mixture of $\ket{0}\ket{\downarrow\uparrow}$ and $\ket{0}\ket{\uparrow\downarrow}$. This is due to the assumption \eqref{eq:inequality-tau} which underpins the optical master equation \eqref{eq:master-mix}. This means we have implicitly included the fast Rabi oscillations between $\ket{\downarrow\uparrow}$ and $\ket{\uparrow\downarrow}$ in the $\mathcal{T}_+$ subspace while the system is waiting for the decay. In contrast, in the $\mathcal{T}_0$ subspace the inherent dynamics is much slower so that the final result depends on the relative magnitudes of $\delta_0$ and $\tau$.

Perhaps surprisingly, decay due to the spontaneous emission of a photon does not act as a source of decoherence if two conditions are met: (i) the system decays from the subspace spanned by the two states $\ket{T_0\downarrow \uparrow}$ and $\ket{T_0\uparrow\downarrow}$, and (ii) $\tau \delta_0 \ll \pi$, i.e.~ the energetic splitting of these two states is small compared to the inverse natural lifetime. This property can be turned into a powerful feature for suitable molecular systems, which we shall exploit in the following.

\subsubsection*{Dealing with a mixed electronic excitation}
So far, we have assumed that the creation process yields a completely polarized excitation, enabling the simple protocol for the generation of entanglement described in a previous section. Motivated by recent experimental data from a promising candidate molecule \cite{Schaffry.Filidou.ea2010EntanglingRemoteNuclear}, we now analyze the implications of having an initial mixture of the states $\ket{T_0}$, and $\ket{T_{\pm}}$.  Experiments on a $^{13}$C labeled methano-carbon of the diethyl malonate mono-adduct (DEMF) reveal that the population of the electronic excitation in a sample oriented along the $z$-axis are equally distributed between $\ket{T_+}$ and $\ket{T_-}$. In this case the lifetime of the excitation also depends on the state of the excitation, being much shorter for the $\ket{T_0}$ state compared to $\ket{T_{\pm}}$. 

In the following we shall demonstrate how the generic protocol presented earlier can be adapted to accommodate for the properties of the specific system presented in Ref.~\onlinecite{Schaffry.Filidou.ea2010EntanglingRemoteNuclear}. After a short laser pulse for the optical excitation, the subspaces $\mathcal{T}_{0,\pm}$ were found to be populated as follows $p_{-}=0.49, p_{0}=0.02,$ and $p_{+}=0.49$ with associated lifetimes $\tau_{-}=\unit{0.57}{\milli\second}, \tau_0=\unit{0.02}{\milli\second},$ and $\tau_+=\unit{0.57}{\milli\second}$. We shall assume that the nuclear spins are initialised in the state $\ket{\downarrow\uparrow}$. 

The basic idea of putting the system into the $\ket{T_{\pm}}$ states to let the free time evolution generate entanglement, followed by (mostly) switching off the interaction in the $\mathcal{T}_0$ subspace remains unchanged. As before switching between different electronic states is accomplished using microwave pulses that are fast on the timescale of the nuclear spin evolution. The adapted protocol proceeds in two stages following the optical excitation. First, we let the desired entanglement build up in the $\mathcal{T}_+$ subspace by waiting for the time $t=\frac{\pi}{4a_+}$. We then swap the populations of $\ket{T_0}$ and $\ket{T_+}$ and wait until the entangled populations have decayed. The difference in decay rates $1/\tau_0 \gg 1/ \tau_{\pm}$ means that the population of $\ket{T_-}$ largely survives once the majority of $\ket{T_0}$ has decayed to the ground state. Second, population in the $\mathcal{T}_-$ subspace is maximally entangled at times that are odd integer multiples of $\frac{\pi}{4a_-}$. We pick the first such point of time after the $\ket{T_0}$ has emptied out and apply another microwave $\pi$-pulse to swap the populations of $\ket{T_0}$ and $\ket{T_-}$. Once more, $\ket{T_0}$ will quickly drain into the ground state, meaning there is now no more excited population left. Ideally, we are left with an almost fully entangled nuclear spin state. 

\begin{figure}[h!]
  \centering
  \includegraphics[scale=1]{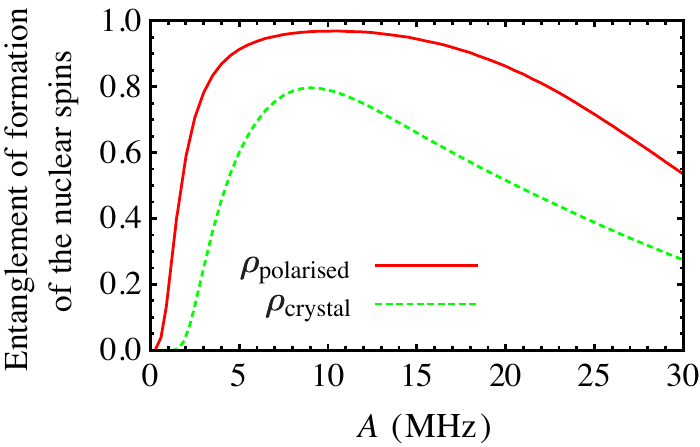}  
 \caption{Entanglement of formation of the nuclear spins after applying our protocols as described in the text for different initial polarizations of the excitation: $\rho_{\text{polarized}} =|T_0\rangle\langle T_0|$ (solid) and $\rho_{\text{crystal}} = 0.49 |T_-\rangle\langle T_-| + 0.02 |T_0\rangle\langle T_0| + 0.49 |T_+\rangle\langle T_+|$ (dashed). For the first case we used the simple protocol described first in this section; the switching time used is $\frac{\pi}{4a_+}$. In the second case we use the enhanced protocol described at the end of this section; switching times here are $\frac{\pi}{4a_+}$ and $\frac{3\pi}{4a_-}$. The nuclear spins for both curves are assumed to be initially in the state $|\downarrow\uparrow\rangle$; for the parameters we use the values found in a recent characterization experiment \cite{Schaffry.Filidou.ea2010EntanglingRemoteNuclear} $D=\unit{-296}{\mega\hertz}$, $\omega_e =\unit{9.6}{\giga\hertz}$, $\omega_n =\unit{3.7}{\mega\hertz}$, $\tau_{-}=\unit{0.57}{\milli\second},$ $\tau_0=\unit{0.02}{\milli\second},$ and $\tau_+=\unit{0.57}{\milli\second}$.}
\label{fig:EntanglementFormation}
\end{figure}

However, the success of the above described protocol is predicated on the coupling strength $A$ of the nuclear spins to the excitation. In particular, for a very small coupling strength $A$ it takes a long time to entangle the nuclear spins $t \approx \frac{\pi\omega_e}{4A^2}$, so that there may be a substantial probability of the optical decay having occurred before the nuclear spins can become properly entangled. In this case the left-hand inequality of equation \eqref{eq:inequality-tau} is violated. On the other hand, if the coupling strength $A$ assumes very large values, then the right-hand side of the inequality \eqref{eq:inequality-tau} is violated. In the latter case the photons resulting from the transitions $\ket{eT_0}\frac{1}{\sqrt{2}}\bigl(\mp\ket{\downarrow \uparrow}+\ket{\uparrow \downarrow}\bigr) \rightarrow \ket{0}\frac{1}{\sqrt{2}}\bigl(\mp\ket{\downarrow \uparrow}+\ket{\uparrow \downarrow}\bigr)$  become distinguishable and the decay is hence no longer coherence preserving. In Fig.~\ref{fig:EntanglementFormation} we regard the hyperfine coupling $A$ as a tunable parameter and plot the entanglement of formation \cite{Wootters1998EntanglementofFormation} of the final state of the two nuclear spins. We consider two different initial states for the mediator spin: a completely polarized state and the mixed state reported in Ref.~\cite{Schaffry.Filidou.ea2010EntanglingRemoteNuclear}.

\subsubsection*{Robustness to imperfections in the symmetry of the system}

The previously described protocol for the controlled generation of entanglement assumes a perfectly symmetric system. In the following we will analyze the degree of imperfection in the symmetry that may be tolerated. We already know that the crossover between the symmetric and the asymmetric case is not entirely abrupt. In Section \ref{sec:cross-betw-symm} we have found expressions for the eigenstates and eigenvalues that can fully interpolate between the symmetric and the asymmetric case. In this crossover case, the effective Hamiltonian still consists of three distinct blocks each corresponding to the spin state of the excitation $\ket{T_j}$ ($j=-,0,+$), and matrix elements connecting the blocks are negligible due to the large energy difference $\omega_e \gg \abs{\omega_n}$. This allows us to assign a separate entangling power $\widetilde{e}_j$ to the effective Hamiltonians describing each of the blocks. 

\begin{figure}[ht!]
  \centering
\includegraphics{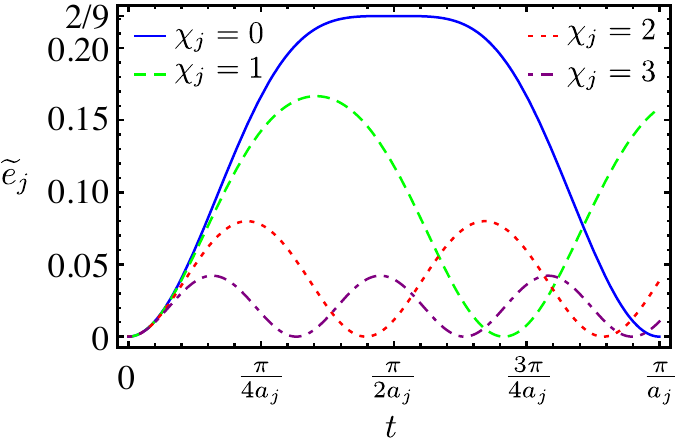}    
  \caption{(Color online) Entangling power $\widetilde{e}_j$ of the free time evolution for an asymmetric system, when the excitation is in the state $|T_j\rangle$ for different strength of the asymmetry $\chi_j$.}
 \label{fig:entangling-power-cross}
\end{figure}

\begin{figure}[ht!]
  \centering
\includegraphics{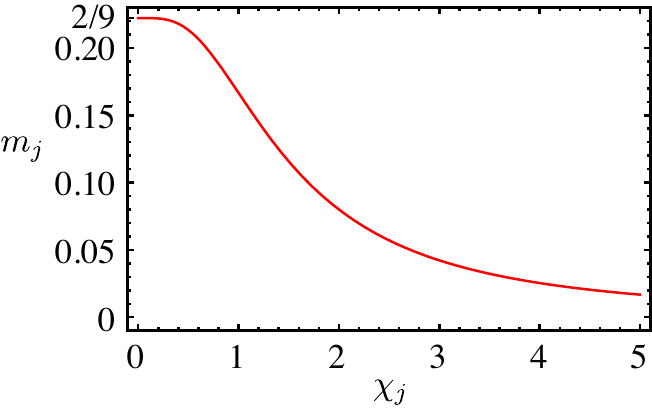}    
  \caption{(Color online) Maximally attainable entangling power $m_j$ of the free time evolution, when the excitation is in the state $|T_{j}\rangle$ with respect to the strength of the asymmetry $\chi_j$.}
  \label{fig:crossover-maxent}
\end{figure}

We now give an easily provable lemma which will enable us to write the relevant analytic expressions of the entangling powers $\widetilde{e}_j$ for all three subspaces: 
\begin{lem}
\label{lem:entangling-power}
Let $H$ be the time-independent Hamiltonian of two spin $\frac{1}{2}$ particles with the following two properties:
\begin{enumerate}
\item $\ket{E_1}=\ket{\downarrow\downarrow}, \ket{E_2}=-a\ket{\downarrow \uparrow} + b \ket{\uparrow \downarrow},\ket{E_3}=b\ket{\downarrow \uparrow} + a \ket{\uparrow \downarrow},$ and $\ket{E_4}=\ket{\uparrow\uparrow}$, with $\abs{a}^2 + \abs{b}^2=1$ and $a,b \in \mathbb{R}$ are eigenvectors of $H$.
\item The eigenenergies  of $H$ satisfy $E_1-E_2-E_3+E_4=0$, 
\end{enumerate}
then the entangling power of $U(t) = \exp(-i H t)$ is given by
\begin{multline}
  \label{eq:ent-general}
  e(b,\beta) = \frac{16}{9}  
(b^2-b^4)\sin^2 \bigl( \tfrac{\beta}{2} \bigr) - \frac{32}{9} (b^2-b^4)^2\sin^4\bigl( \tfrac{\beta}{2} \bigr) \;,
\end{multline}
where $\beta= \abs{E_3 -E_2}t$. In addition we have $e(a,\beta)=e(b,\beta)$.
\end{lem}
Applying the above Lemma to our effective Hamiltonians for the crossover case gives three expressions $ \widetilde{e}_{j} = e(c_j,\beta_j)$ with 
\begin{equation}
 \beta_j = 2t \sqrt{a_j^2 + f_j^2} \quad \text{and} \quad c_j^2 = \frac{1}{2}\left(1 + \frac{f_j}{\sqrt{a_j^2+f_j^2} } \right) \quad.
\end{equation}
It is easy to see that the parameter $a_j$ now directly competes with the strength of the perturbation $f_j$ in the expression for the entangling power. With the following measure for the asymmetry 
\begin{equation}
 \chi_j = \abs{\frac{f_j}{a_j}}
\end{equation}
we can write the entangling power compactly as 
\begin{equation}
\widetilde{e}_j = \frac{\left(3+4 \chi_j^2+\cos\left(2 a_j t
      \sqrt{1+\chi_j^2}\right) \right) \sin\left(a_j t
    \sqrt{1+\chi_j^2}\right)^2}{9 \left(1+\chi_j^2\right)^2} \quad.
\end{equation}
We plot $ \widetilde{e}_{j}$  for different asymmetries $\chi$  in Fig.~\ref{fig:entangling-power-cross}. 
Not surprisingly, the entangling power decreases with increasing asymmetry $\chi_j$. The maximum of the entangling power, which is achieved for $\beta_j=k\pi$ where $k$ is an odd integer, is given by
\begin{equation}
\label{eq:maxEnt}
  m_j = \frac{16}{9} (c_j^2 -c_j^4) - \frac{32}{9}(c_j^2 -c_j^4)^2 =\frac{2}{9} \frac{ 1+2 \chi_j^2}{(1+\chi_j^2)^2} \quad;
\end{equation}
this expression is plotted in Fig.~\ref{fig:crossover-maxent}.
\begin{figure}
  \centering
  \includegraphics{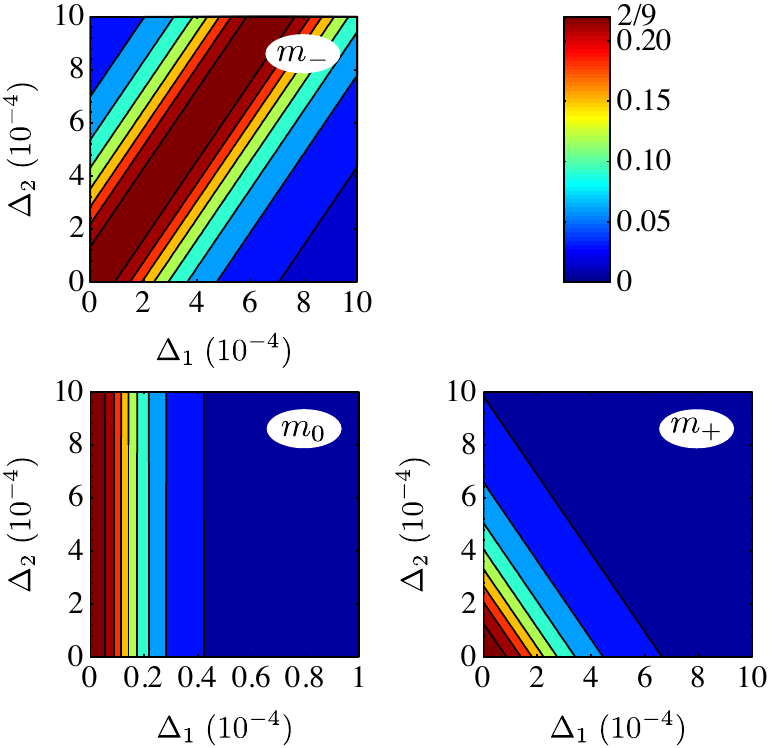}
  \caption{Maximal attainable entangling power $m_-, m_0,$ and $m_+$ as a function of the asymmetry in the Zeeman splitting $\Delta_1=\frac{\omega_{n'}-\omega_n}{\omega_n}$ and hyperfine coupling asymmetry $\Delta_2=\frac{A'-A}{A}$. Other parameters are: $A=\unit{2.5}{\mega\hertz}$, $\omega_n=\unit{3.7}{\mega\hertz}$, $D=\unit{-296}{\mega\hertz}$, $\omega_e=\unit{9.6}{\giga\hertz}$. The entangling  power $m_j$ only reaches its maximum limit of 2/9 for certain values of $\Delta_1$ and $\Delta_2$. See main text for a more detailed discussion.}
  \label{fig:contourPlot}
\end{figure}
There is no significant reduction in the achievable entanglement power as long as $\chi_j < 1/2$, but the maximum drops quickly outside this regime. We note that for equal hyperfine couplings ($\Delta_2 = 0$) but unequal nuclear gyromagnetic ratios
\begin{equation}
\frac{\chi_0}{\chi_{\pm}} = \frac{a_{\pm}}{\abs{a_0}} \approx \frac{\omega_e}{2\abs{D}} \gg 1  \quad,
\end{equation}
meaning that the $\mathcal{T}_0$ subspace's entangling power $\widetilde{e}_0$ is much more affected by the asymmetry than $\widetilde{e}_{\pm}$. Fortunately, the dynamics in this subspace is also the slowest, so that it can still conveniently serve as a shelf for entangled states generated in $\mathcal{T}_+$ or $\mathcal{T}_-$ until the optical excitation has been de-excited or decayed. Importantly, Eq.~\eqref{eq:maxEnt} implies that our scheme is robust against the small deviations from a perfectly symmetrical system which one might expect in real-world experiments. Further, intentionally introduced small differences between the frequencies $\omega_n$ and $\omega_{n'}$ (e.g.~a chemical shift caused by different surrounding environments) may actually be useful for individual control and tomography of the nuclear spins, while a high-fidelity entangling operation is still possible.

So far we have discussed the behaviour of the entangling power in terms of the parameter $\chi_j$. While the dependence of the $m_j$ on $\chi_j$ is universal across the three subspaces $\mathcal{T}_j$, we obtain qualitatively different results when considering plots that are based directly on the $\Delta_{1/2}$ asymmetry parameters. Fig.~\ref{fig:contourPlot} shows the entangling power as a function of $\Delta_1$ and $\Delta_2$. We see that the behaviour is indeed quite different in each of the three subspaces: In the $\mathcal{T}_-$ subspace we obtain a ridge along which an asymmetry between the nuclear Zeeman splittings and the hyperfine coupling constants completely cancels out, meaning a perfect operation is possible even for a system that is quite far removed from being symmetric. In contrast, the asymmetries add up in the  $\mathcal{T}_+$ subspace, so that the error tolerance is much reduced in this case. Finally, the  $\mathcal{T}_0$ subspace is only sensitive to $\Delta_1$, i.e.~the difference in the nuclear Zeeman splittings without any dependence on the hyperfine constants (see Eq.~\eqref{eq:defOfF}). 

The entangling power vanishes completely in the limit of an entirely asymmetric system (which we take to be defined by the inequalities \eqref{eq:cond} and \eqref{eq:condB}). In this case $\chi_j \gg 1$, and the eigenstates consequently coincide with the computational basis states, and the eigenvalues are such that the free time evolution no longer generates any entanglement (i.e. requirement 2 of the Lemma is satisfied). We shall analyze this situation in the following section.

\subsection{Control methods for the asymmetric system}
\label{sec:asymm-case:-micro}

\begin{figure}[hbt]
  \centering
  \includegraphics[scale=1]{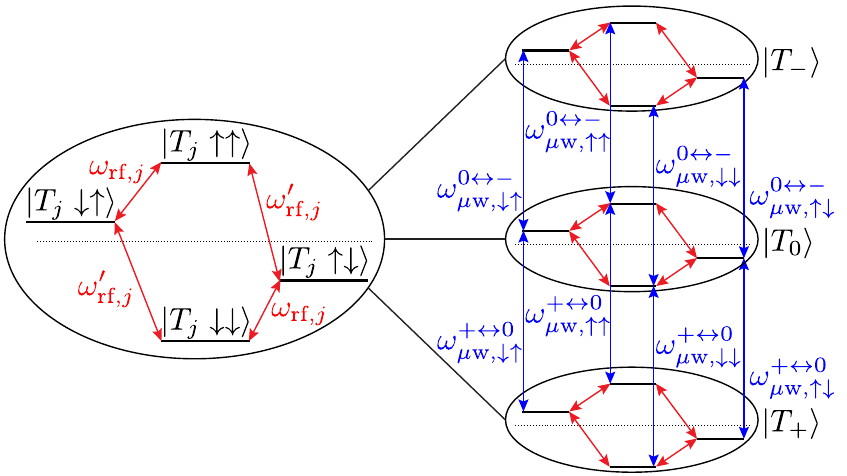}  
  \caption{(Color online) Microwave (blue) and radio-frequency transitions (red) of the asymmetric system. The computational basis states are eigenstates of the system Hamiltonian (assuming Eqs.~\eqref{eq:cond} and \eqref{eq:condB} are fulfilled).}
  \label{fig:transitions}
\end{figure}

For an asymmetric system we cannot rely on the system's free time evolution for the generation of entanglement; this is a direct consequence of the Hamiltonian being decomposable into local Hamiltonians (see Eq.~\eqref{eq:AsymmetricHamiltonian}). Therefore, we need to apply a suitable sequence of radio-frequency and microwave control pulses to accomplish our aim of creating an entangled nuclear spin state. Hence, we proceed by analysing the dipole-allowed transitions of the asymmetric system (see Fig.~\ref{fig:transitions}). 

The asymmetric system possesses six (different) nuclear spin transitions on the radio-frequency scale, one per nuclear spin for each of the three spin states of the excitation. Referring back to section \ref{sec:asymmetric-case} we obtain these from the second order eigenenergies:
\begin{align}
  \omega_{\text{rf},j} &= \abs{-j A - \omega_n - a_- ( \delta_{j,-} + \delta_{j,0} ) - a_+ ( \delta_{j,0} + \delta_{j,+} ) },\\ 
  \omega'_{\text{rf},j} &= \abs{-j A' - \omega_{n'} - a'_- ( \delta_{j,-} + \delta_{j,0} ) - a'_+ ( \delta_{j,0} + \delta_{j,+} )}, 
\end{align}
where $\delta_{k,l}$ is the Kronecker delta, and as before $j$ indexes the spin state of the excitation $\ket{T_j}$. Further, $\omega_{\text{rf},j}$ denotes the transition frequency of the first nuclear spin and $\omega'_{\text{rf},j}$ the transition frequency of the second nuclear spin. In general all six of these frequencies may be distinct.

Conversely, the spin state of the electronic excitation can be flipped conditional on the nuclear spin state using a suitable microwave pulse. With four nuclear spin states and two excitation spin transitions, this gives a total of eight microwave frequencies taking the excitation from $\ket{T_j}$ to $\ket{T_{j'}}$ with $(j,j')=(+,0)$ or $(j,j')=(0,-)$ or \textit{vice versa}. These are:
\begin{align}
\label{eq:mw1}
  \omega_{\mu\text{w,}\downarrow\downarrow/ \uparrow\uparrow}^{j \leftrightarrow j'} &= \omega_e + (-1)^j D \pm \frac{1}{2} (A +A')\\
  \omega_{\mu\text{w,}\downarrow\uparrow/ \uparrow\downarrow}^{j \leftrightarrow j'} &= \omega_e + (-1)^j D \pm \frac{1}{2} (A -A') \quad.  \label{eq:mw2}
\end{align}
Here we have neglected second order perturbation theory shifts proportional to $a_j$ and $a'_j$, as these are typically very small when compared to $A$ and $D$.

Several possibilities exist for exploiting this rich transition spectrum in order to realise an entangling operation. In the following we shall discuss three methods in more detail: a single microwave $2\pi$-pulse, a pulse sequence of radio- and microwave pulses, and finally an adiabatic following method.
 
\subsubsection{CPHASE-gate through a selective $2\pi$-pulse}
\label{sec:pi-pulse}

Simply applying a selective $2\pi$-pulse with the frequency of any of the microwave transitions given in Eq.~\eqref{eq:mw1} and \eqref{eq:mw2} naturally implements a CPHASE-gate by imparting a phase of $e^{i \pi} = -1$ to only one of the four nuclear spin states \cite{Nielsen.Chuang2000QuantumComputationand}.

If the lifetime of the transition was infinite (and in the absence of other spin dephasing mechanisms), the $2\pi$-pulse could be made perfectly selective, achieved by a pulse that is long in the time domain and accordingly spectrally narrow in the frequency domain \cite{Bernstein.King.ea2004HandbookofMRI}. However, in practice the optical lifetime will be finite and this may limit the selectivity and thus the amount of entanglement that can be achieved. The challenge is to find the right balance between a fast pulse that is only partially selective, and a slow pulse during which the system suffers from the decoherence induced by the decay. We proceed by analysing the trade-off that arises from these constraints in the following.

Suppose we are given an initial state that is an equal superposition of the computational basis states and a fully polarized state of the excitation,
\begin{equation}
\label{eq:plusplus}
 \ket{\psi_{\text{initial}}} = \frac{1}{2}\ket{T_0} ( \ket{\downarrow\downarrow} + \ket{\downarrow\uparrow} + \ket{\uparrow\downarrow} + \ket{\uparrow\uparrow}) \quad,
\end{equation}
and then apply a $2\pi$-pulse microwave pulse with power $\Omega_0$ and with a frequency $\omega_D$ corresponding to the energy difference between the levels $\ket{T_+\uparrow\uparrow}$ and $\ket{T_0\uparrow\uparrow}$. To describe the dynamics of the excited system we use either the effective asymmetric or the more general effective crossover Hamiltonian, whichever is more appropriate for the precise combination of system parameters in question. In particular, $H_{\text{asym,eff},j}$ is adequate whenever the eigenvectors closely coincide with the computational basis states, whereas $H_{\text{co,eff},j}$ is used otherwise. We apply the following criterion for discriminating between the two cases:
\begin{equation}
H_{\text{eff},j}  =\begin{cases}
   H_{\text{asym,eff},j}  & \abs{\alpha_{j,2,1}}^2 \leq 0.001 \text{ or }\abs{\alpha_{j,2,1}}^2 \geq 0.999 \quad, \\
  H_{\text{co,eff},j}    &    0.001< \abs{\alpha_{j,2,1}}^2< 0.999 \quad,
  \end{cases}
\end{equation}
with $\alpha_{j,2,1}$ as defined in Eq.~\eqref{eq:alpha}. On top of the effective system Hamiltonian $H_{\text{eff}}$ we also need to model the microwave pulse (in the usual RWA), so that the total Hamiltonian during the pulse is given by:
\begin{multline}
  H_{\mu\text{w}} =\ket{0}(-\omega_n S_{z,n}- \omega_{n'} S_{z,n'})\bra{0} + \ket{e} H_{\text{eff}} \bra{e}\\
           +\ket{e}(\Omega_0 S_{x,e}-\omega_D S_{z,e})\bra{e}  \quad.
\end{multline}
As this Hamiltonian is time-independent we can use a quantum optical master equation like the one defined in Eq.~\eqref{eq:master-mix} to model the decay of the excitation in the interaction picture. The transition operators $A(\omega)$ are as defined by Eq.~\eqref{eq:transitionOperator}, with appropriate projectors onto the eigenspaces of $H_{\mu\text{w}}$. In our calculations we perform the RWA in the master equation (remember that this RWA is different and independent from the RWA for the driving) whenever the difference of two frequencies differs by more than $30 \tau^{-1}$, where $\tau$ denotes the lifetime of the excitation.  

While optical decay during the application of the pulse may preserve some coherence between the nuclear spin states for specific parameter combinations, this is no longer the case once the pulse has finished: the decay to the ground state for an asymmetric system in the absence of microwave driving invariably destroys the nuclear coherences. Therefore, we must use a different approach for taking the system back to its ground state. The first possibility is de-excitation using a resonant optical $\pi$-pulse. Another option would be to significantly `speed up' the decay process, e.g. by exciting the system into a different metastable excited state which is known to quickly decay to the ground state. Provided the lifetime of this metastable state is short enough, the wave function of the emitted photon does not carry information about the nuclear spin state, so that the nuclear spin coherence will be preserved. We shall assume that such a coherence preserving de-excitation can be accomplished. 

As mentioned above, the system is susceptible to potentially harmful decay events during the application of the $2\pi$-pulse, such that the final nuclear spin state $\rho_{\text{nuc}}$ is a mixture of population that has  spontaneously decayed and the remaining population to which the control sequence has been fully applied.  Since we are now dealing with mixed states $\rho_{\text{nuc}}$, the entangling power is no longer a suitable measure for the quality of our operation, and as before we employ the entanglement of formation \cite{Wootters1998EntanglementofFormation} as an alternative benchmark. 

\begin{figure}[hbt]
  \centering
 \includegraphics{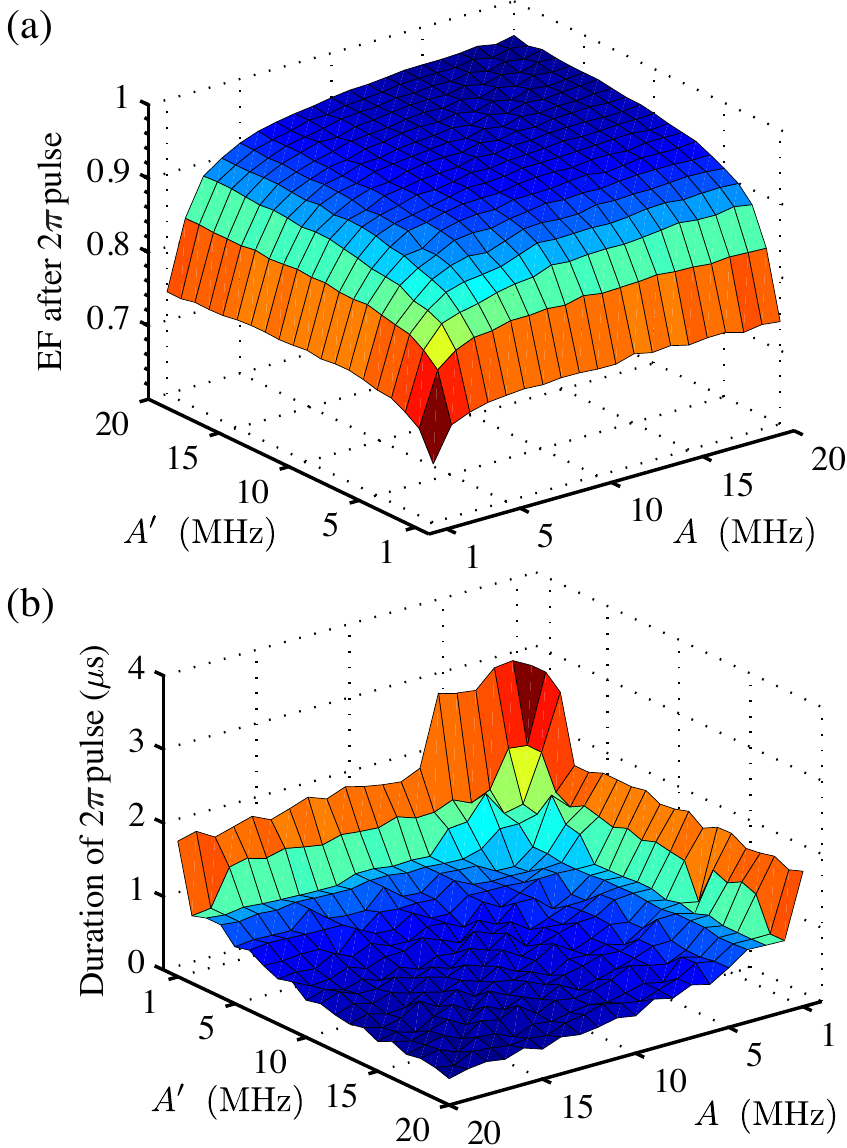}
 \caption{(Color online) (a) Maximized entanglement of formation of he nuclear spins after the $2\pi$-pulse with respect to the two hyperfine coupling strengths $A$ and $A'$. (b) Optimal duration $t^*=\frac{2\pi}{\sqrt{2}\Omega_0^*}$ of the microwave $2\pi$-pulse with respect to the two hyperfine coupling strengths $A$ and $A'$. We use the illustrative lifetime  $\tau=\unit{10}{\micro\second}$ while the other parameters used in this plot are motivated by Ref.~\cite{Filidou.others}: $D=\unit{-320}{\mega\hertz},\omega_e=\unit{9.7}{\giga\hertz},\omega_n=\unit{5.97}{\mega\hertz},\omega_{n'}=\unit{14.74}{\mega\hertz}$.}
  \label{fig:EF2pi}
\end{figure}

Assuming a simple top hat pulse profile in the time domain, the pulse duration for a $2 \pi$-pulse is $t=\frac{2\pi}{\sqrt{2}\Omega_0}$ where $\Omega_0$ is the applied microwave power. For optimal performance the right balance must be found between pulse selectivity and duration for each combination of system parameters and lifetime $\tau$.  We thus  maximize the achievable entanglement of formation by varying $\Omega_0$ to obtain
\begin{equation}
\label{eq:maxEFopt}
\text{EF}^* = \max_{\Omega_0} \text{EF}(\rho_{\text{nuc}})  \quad.
\end{equation}
As an example we choose $\tau=\unit{10}{\micro\second}$ to plot the quantity $\text{EF}^*$ in Fig.~\ref{fig:EF2pi} as a function of $A$ and $A'$. Larger hyperfine coupling constants allow a faster selective pulse, and the optimized entanglement of formation of $\rho_{\text{nuc}}$ hence increases with $A$ and $A'$. Remarkably, even for a lifetime as short as \unit{10}{\micro\second}, a high entanglement of formation can be obtained with only moderate hyperfine coupling strengths. 

Finally, we note that the currently presented protocol also works for the symmetric system, where the entanglement operation then only takes a time $t=\frac{\pi}{2a_{\pm}}\approx \frac{\pi \omega_e}{A^2}$, which is much faster than our protocol discussed in Section \ref{sec:symmtimeev}. For a short optical lifetime this approach may thus be advantageous assuming a spectrally narrow highly selective $2 \pi$ pulse can be implemented.

\subsubsection{Combined microwave and radio-frequency pulse sequence}
\label{sec:shelving-pulse}
\begin{figure}[hbt]
  \centering
  \includegraphics[scale=1]{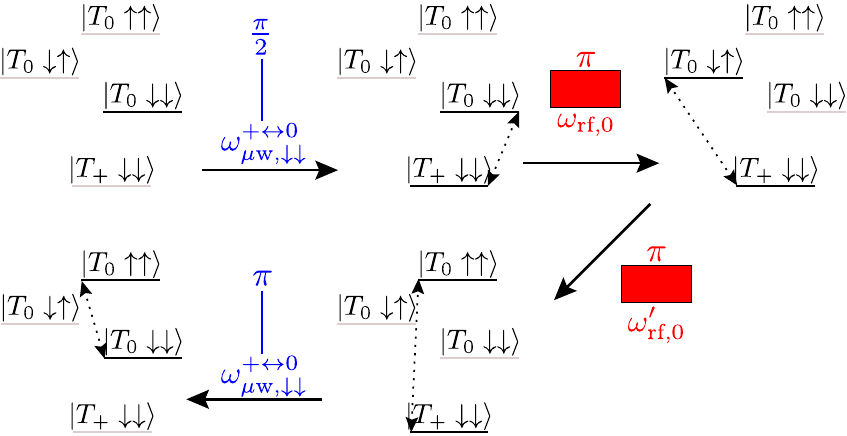}
  \caption{(Color online) Pulse sequence for the creation of the Bell state $\frac{1}{\sqrt{2}}\bigl(|T_0\downarrow \downarrow\rangle+|T_0\uparrow\uparrow\rangle \bigr)$. The black energy levels are populated and the grey levels are empty. The dotted arrows denote a coherence between the adjacent energy levels. Only relevant spin levels are shown.}
  \label{fig:shelving}
\end{figure}

In the previous section we have discussed a simple implementation of the CPHASE-gate that is maximally entangling for suitable system parameters. However, other methods for creating maximally entangled states also exist, and these may be more advantageous if a particular final state is required. Here we shall briefly discuss an alternative control method that employs a sequence of microwave and radio frequency pulses instead of a single microwave pulse. Let us assume we have the initial spin state $\ket{T_0\downarrow \downarrow}$ and would like to create the entangled Bell state $\frac{1}{\sqrt{2}}\bigl(\ket{T_0\downarrow \downarrow}+\ket{T_0\uparrow\uparrow} \bigr)$. It is impossible to perform this operation with only radio-frequency pulses. However, we can achieve our aim by `shelving' parts of the population in one of the other electronic spin states. An example of how this approach works in detail is depicted in Fig.~\ref{fig:shelving}. 

Of course, the microwave and radio-frequency pulses also need to be sufficiently selective for this approach, which imposes a minimal overall pulse sequence duration. Once more, we refrain from discussing  sophisticated pulse shaping techniques \cite{Bernstein.King.ea2004HandbookofMRI}, instead considering considering Rabi's instructive formula for the transition probability of a driven two level system 
\begin{equation}
  \mathcal{P}_{1\rightarrow2}(t)= \frac{\Omega^2}{\Omega^2+\Delta^2} \sin^2(\sqrt{\Omega^2+\Delta^2}t) \quad, 
\end{equation}
where $\Omega$ denotes the strength of the pulse and $\Delta$ the detuning. We apply this  to the closest neighbouring transitions of the (resonantly driven) target transition and demand that $\Omega^2 / (\Omega^2 + \Delta^2) \ll 1$ for all neighbouring transitions, allowing a rough estimation of the duration required for each pulse in the sequence.

We take the example of a recent characterization experiment on an asymmetric system (phosphine oxide fullerene DMFPH) \cite{Filidou.others}, where the two nuclear spins of interest are provided by a hydrogen and phosphorus atom in the functional group attached to the fullerene. The hyperfine coupling and the ZFS were measured as
\begin{equation}
  A\approx\unit{6}{\mega\hertz},\quad A'\approx \unit{11}{\mega\hertz}, \text{ and } D\approx\unit{-320}{\mega\hertz} 
\end{equation}
at an external magnetic field of $B=\unit{0.346}{\tesla}$. Setting an upper bound of $0.01$ for the unwanted transition probabilities, the four required pulses can all be applied in less than a \unit{1}{\micro\second} for typical nuclear gyromagnetic ratios. The duration of the pulse sequence is thus short compared to expected optical lifetimes in candidate molecules, so that a high fidelity entangling operations using this protocol should be feasible. However, we refrain from performing a more detailed analysis as in the previous section since the results are similar and little additional insight is gained.

\subsubsection{Implementation of a CPHASE operation with adiabatic following}
\label{sec:adiabatic-scheme}

The last method discussed in this paper for creating entanglement in our system relies on the adiabatic following of system eigenstates, similar to the protocol described in Refs.~\cite{Gauger.Rohde.ea2008Strategiesentanglingremote,Gauger.ea2008.RobustApproach}. Here, it is implemented by slowly modulating the intensity of a microwave pulse that is close to resonance with one or several of the microwave transitions of the excitation spin. Prior to the application of the pulse, the (asymmetric) system is prepared to be in a superposition of computational basis states as follows:
\begin{equation}
\label{eq:superpos}
 \ket{\psi_{\text{initial}}} = \ket{T_0} (a_1 \ket{\downarrow\downarrow} + a_2 \ket{\downarrow\uparrow} + a_3\ket{\uparrow\downarrow} + a_4\ket{\uparrow\uparrow}) \quad,
\end{equation}
with normalisation $\quad \sum_{i=1}^4\abs{a_i}^2 =1$. Starting from this state, the microwave power is then varied such that adiabatic following of instantaneous eigenstates occurs. Once the power is decreased again all population returns back to the computational basis. During the pulse, the eigenstates are energetically shifted and thus pick up a dynamic phase. However, the precise shifts of the states differ due to the the hyperfine coupling, so that in general each of the four states acquire a different dynamic phase. This gives rise to an overall combination that can be non-trivial and entangling \cite{Gauger.ea2008.RobustApproach}.

Consider applying a microwave field with frequency $\omega_D$ whose power envelope is changed gradually following a Gaussian function $\Omega(t) = \Omega_0 \exp[-(t/\sigma)^2 ]$. We apply this Gaussian microwave pulse from an initial time $t=-3\sigma$ to $t=3\sigma$ and choose the frequency $\omega_D$ in such a way that the pulse is off-resonant with all microwave transition in the system. The diagonal form of the system Hamiltonian $H_{\text{asym,eff}}$ permits a description of the dynamics of each state in the superposition of Eq.~\eqref{eq:superpos} with a Hamiltonian connecting all three excitation states of the form
\begin{equation}
   \mathcal{H}_i(t)=\bra{i}\mathcal{H}\ket{i}=
\begin{pmatrix}
  \Delta_{i,1} & \frac{\Omega(t)}{\sqrt{2}} & 0 \\
  \frac{\Omega(t)}{\sqrt{2}}  & 0 &   \frac{\Omega(t)}{\sqrt{2}}  \\
  0 & \frac{\Omega(t)}{\sqrt{2}}    & \Delta_{i,2} 
\end{pmatrix}
\end{equation}
written in the basis $\{\ket{T_- i},\ket{T_{0} i},\ket{T_+ i}\}$ for $i=\{\downarrow\downarrow,\downarrow\uparrow,\uparrow\downarrow,\uparrow\uparrow\nolinebreak\}$ with detunings $\Delta_{i,1}=\omega_{\mu\text{w,}i}^{0 \leftrightarrow -}-\omega_D$ and $\Delta_{i,2}=\omega_{\mu\text{w,}i}^{+ \leftrightarrow 0}-\omega_D$. Note that the usual RWA has been performed. Choosing $\omega_D$ such that for each $i$ 
\begin{equation}
\label{eq:perturbation-cond}
\frac{\Omega(t)}{\sqrt{2}}\ll \Delta_{2,i}   \quad
\end{equation}
then enables us to write an approximate Hamiltonian for each of the nuclear spin states (valid to first order in perturbation theory) as
\begin{equation}
   \mathcal{H}_{i,\text{app}}(t)=
\begin{pmatrix}
  \Delta_{i,1} & \frac{\Omega(t)}{\sqrt{2}} & 0 \\
  \frac{\Omega(t)}{\sqrt{2}}  & 0 & 0   \\
  0 & 0  & \Delta_{i,2} 
\end{pmatrix} \quad.
\end{equation}
To achieve adiabatic following the eigenenergies need to be varied slowly to suppress Landau Zener transitions between different eigenstates. Following Ref.~\onlinecite{Gauger.Rohde.ea2008Strategiesentanglingremote} this can be accomplished under the following conditions:
\begin{equation}
  \label{eq:AdiabCond}
  \frac{\Omega_0}{\Delta_{1,i}^2}\ll \sigma  \quad \text{for } i=\{\downarrow\downarrow,\downarrow\uparrow,\uparrow\downarrow,\uparrow\uparrow \} \quad.
\end{equation}
As we have mentioned above, the net effect achieved by the adiabatic pulse is the acquisition of a phase $\theta_i$ for each of the nuclear spin states. The `right' combination of control parameters $\sigma, \Omega_0,$ and $\Delta_{1,i}$ gives rise to an operation that is locally equivalent to a CPHASE-gate if the following condition is met \cite{Gauger.Rohde.ea2008Strategiesentanglingremote}:
\begin{equation}
  \label{eq:sum-pi}
  \pi = \theta_1 - \theta_2 - \theta_3 + \theta_4 \quad.
\end{equation}
In terms of the evolution of the nuclear spins, the state $\ket{\psi_{\text{initial}}}$ has then evolved to 
\begin{equation}
  \ket{T_0} (a_1 e^{i\theta_1} \ket{\downarrow\downarrow} + a_2 e^{i \theta_2} \ket{\downarrow\uparrow} + a_3 e^{i \theta_3} \ket{\uparrow\downarrow} + a_4 e^{i \theta_4}\ket{\uparrow\uparrow}), 
\end{equation}
corresponding to a local phase on each nuclear spin in addition to the application of a CPHASE-gate.

We shall now address the question of how the optimal combination of control parameters may be found. The dynamical phase that is acquired by a state during the pulse is directly determined by the eigenenergy of the state $\ket{T_0 i}$, yielding
\begin{align}
   \theta_i  &=  -\int_{-3\sigma}^{3 \sigma} \frac{1}{2} \left( \Delta_{i,1} - \sqrt{ \Delta_{i,1}^2 + 2\Omega(t)^2} \right) dt \\
&= \frac{\Omega_0 \sigma}{2} \int_{-3}^3 \sqrt{ \left(\frac{\Delta_{i,1}}{\Omega_0} \right)^2 + 2\exp(-2x^2)} dx \quad .
\end{align}
Imposing the condition~\eqref{eq:sum-pi} and solving for $\sigma=\sigma(\omega_D, \Omega_0)$ yields
\begin{equation}
\sigma = \frac{2\pi}{\Omega_0} \Biggl(  \int_{-3}^3 \sum_{i} \sqrt{ \left(\frac{\Delta_{i,1}}{\Omega_0} \right)^2 + 2\exp(-2x^2) dx}  \Biggr)^{-1}  \quad,
\end{equation}
where the sum is taken over the four nuclear spin eigenstates. To mitigate the effect of decoherence caused by the decay of the excitation we minimize the duration of the pulse under the constraint that the conditions in Eqs.~\eqref{eq:perturbation-cond}, \eqref{eq:AdiabCond}, and \eqref{eq:sum-pi} are fulfilled, thus obtaining an optimal $\sigma^*$.

We incorporate the decoherence caused by the optical decay by modelling the time evolution with the following Schr\"odinger picture master equation
\begin{multline}
  \label{eq:master-rwa-schroedinger}
\frac{d}{dt} \rho(t) = -i[\mathcal{H}_{\text{app}}(\sigma^*),\rho(t)] \\
+ \sum_{\omega}  \Gamma(\omega) \Bigl(2 A(\omega) \rho(t) A(\omega)^{\dagger} -\{ A(\omega)^{\dagger}A(\omega), \tilde{\rho}(t) \}\Bigr) 
\end{multline}
in Lindblad form \cite{Breuer.Petruccione2002TheoryofOpen}. $\Gamma(\omega)$ is as defined by Eq.~\eqref{eq:rates} and the Lindblad operators $A(\omega)$ are determined by \eqref{eq:transitionOperator}, where the projectors project onto the eigenspaces of $H_{\text{asym,eff}}$ instead of $\mathcal{H}_{\text{app}}$. This simplification gives twelve (constant) incoherent decay channels, rather than considering time-dependent Lindblad operators and re-evaluating the validity of the RWA at every instance in time (as would be required for the time-dependent Hamiltonian). Effectively, our approach then overestimates the destructiveness of the optical decay, thus giving a lower bound for the entanglement of formation of the nuclear spins. 

\begin{figure}[hbt]
  \centering
\includegraphics{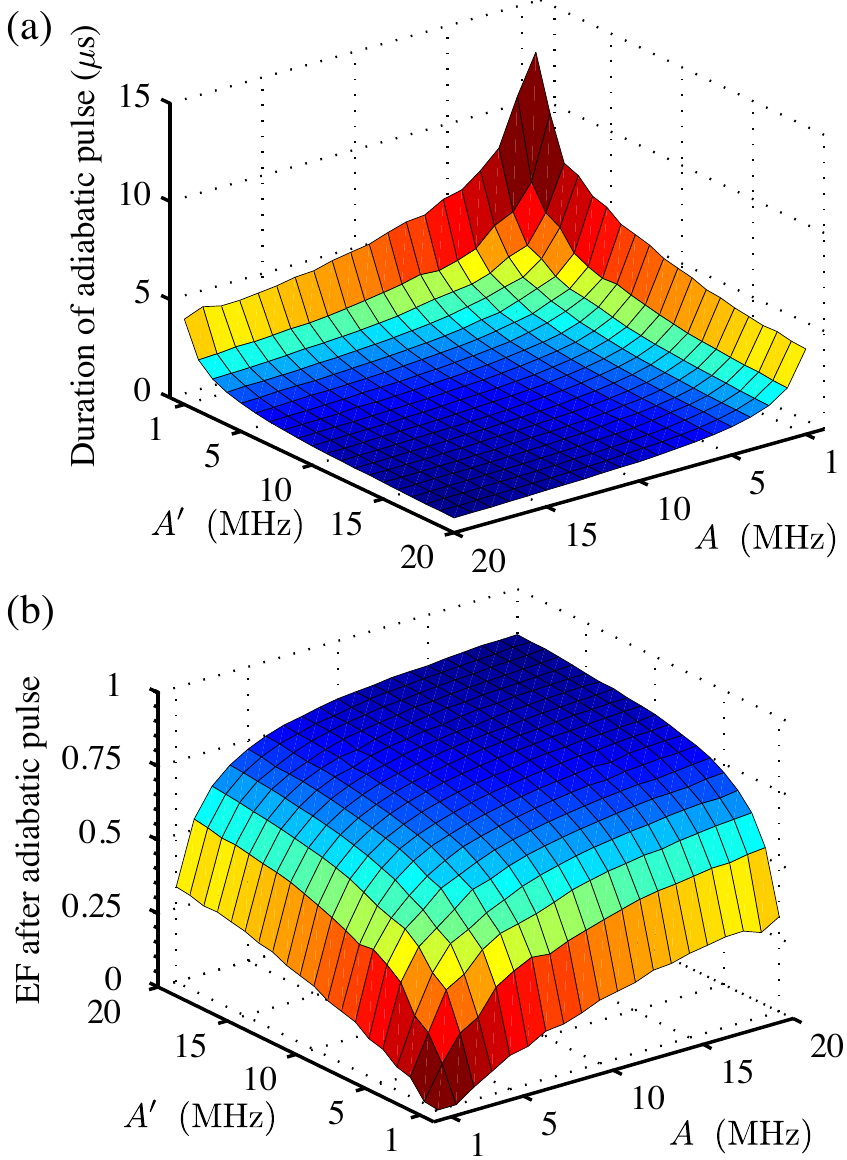}
 \caption{(Color online) (a) (b) Optimal duration of the adiabatic pulse $6\sigma^*$ optimised over $\omega_D$ and $\Omega_0$. (b) Entanglement of formation after applying a Gaussian pulse whose duration is characterised by $\sigma^*$. As in Fig.~\ref{fig:EF2pi} other parameters are: $\tau=\unit{10}{\micro\second}, D=\unit{-320}{\mega\hertz},\omega_e=\unit{9.7}{\giga\hertz},\omega_n=\unit{5.97}{\mega\hertz},\omega_{n'}=\unit{14.74}{\mega\hertz}$.}
 \label{fig:adiabaticPulse}
\end{figure}

Fig.~\ref{fig:adiabaticPulse} shows the results of a simulation that applies such an optimised Gaussian pulse with a pulse duration $\sigma^*$. For weak coupling strengths $A$ and $A'$, this approach achieves a somewhat lower value of the entanglement of formation than the dynamic $2 \pi$-pulse discussed earlier. However, a similarly high fidelity entangling operation is possible for stronger hyperfine coupling. The term `adiabatic following' can invoke the impression that the desired operation will be much slower than a dynamical implementation. It is therefore astounding that our adiabatic pulse only takes about twice as along as the dynamic $2\pi$-pulse. Finally we note that the adiabatic method (where the pulse is applied off-resonantly rather than having to hit a specific resonance) is inherently robust against pulse imperfections. This could be a significant advantage for experiments with ensembles of identical molecules. In this case, static and driving field inhomogeneities will inevitably lead to an over- or under-rotation of some of the  ensemble spins when a  dynamical pulse is applied (leaving the wrong excitation subspace populated), whereas the adiabatic approach ensures that all populations end up in back in the correct spin levels.

\section{Summary}
\label{sec:summary}
In this paper, we have given a detailed analysis showing how a transient optically excited state can be harnessed for the controlled generation of entanglement between two remote nuclear spins. We have identified control methods applicable over a wide range of system parameters and studied their performance with regard to the predominant decoherence mechanism. For the symmetric system consisting of two identical nuclear spins as qubits, the free time evolution is naturally entangling, but the characteristic timescale of the dynamics depends on the state of the excitation and can vary over several orders of magnitude, opening up the possibility of effectively switching the interaction on and off. For an asymmetric system, a different route needs to be taken and we have presented one adiabatic and two dynamic methods for creating entanglement in this case. We have also included a discussion of the crossover regime between the asymmetric and symmetric system to establish the robustness of the symmetric operation.

We have shown that the symmetric control method can be remarkably robust against uncertainty or fluctuations in the coupling constants and nuclear Zeeman splittings. As another advantage of the symmetric system, the system can decay back to the ground state without destroying the nuclear spin coherence. Conversely, for the asymmetric system additional control is required for the de-excitation step, yet it is easier to address the nuclear spins individually for single qubit operations, initialisation and read-out. 

Interestingly, the active control methods proposed for the asymmetric system are much faster than waiting for the free time evolution in the symmetric case, and they can also be applied to the symmetric system if a short optical lifetime makes this approach advantageous. Finally, we note that the adiabatic control method is intrinsically more robust against control pulse and  static field inhomogeneities, making it uniquely suitable for experiments with ensembles of identical systems. Astonishingly, the time required for such an adiabatic operation is only about twice as long as for its dynamical counterpart.

\begin{acknowledgments}
We thank Vasileia Filidou, Simon C. Benjamin and John J. L. Morton for fruitful discussions. This work was supported by the Marie Curie Early Stage Training network QIPEST (MESTCT-2005-020505), EPSRC through QIP IRC (GR/S82176/01 and GR/S15808/01), the DAAD, Linacre College, and the Royal Society.
\end{acknowledgments}


%

\end{document}